# Development and Characterization of a Diamond-Insulated Graphitic Multi Electrode Array Realized with Ion Beam Lithography

**Federico Picollo** [1] [2]*, **Alfio Battiato** [2] [1], **Emilio Carbone** [3], **Luca Croin** [4] [5], **Emanuele Enrico** [5], **Jacopo Forneris** [2] [1], **Sara Gosso** [3], **Paolo Olivero** [2] [1], **Alberto Pasquarelli** [6] and **Valentina Carabelli** [3]

1. Istituto Nazionale di Fisica Nucleare, sezione di Torino, via P. Giuria 1, 10125 Torino, Italy
2. Physics Department and NIS Center, University of Torino, via P. Giuria 1, 10125 Torino, Italy; E-Mails: battiato@to.infn.it (A.B.); forneris@to.infn.it (J.F.); olivero@to.infn.it (P.O.)
3. Department of Drug Science and Technology and NIS Center, University of Torino, Corso Raffaello 30, 10125 Torino, Italy; E-Mails: valentina.carabelli@unito.it (V.C.); emilio.carbone@unito.it (E.C.); sara.gosso@unito.it (S.G.)
4. Department of Applied Science and Technology - DISAT, Politecnico di Torino, Corso Duca degli Abruzzi 24, 10129 Torino, Italy; E-Mail: luca.croin@polito.it (L.C.)
5. Nanofacility Piemonte, National Institute of Metrologic Research (INRiM), Strada delle Cacce 91, 10135 Torino, Italy; E-Mail: e.enrico@inrim.it (E.E.)
6. Institute of Electron Devices and Circuits, Ulm University, 89069 Ulm; E-Mail: alberto.pasquarelli@uni-ulm.de (A.P.)

* Author to whom correspondence should be addressed; E-Mail: picollo@to.infn.it; Tel.: +39-011-670-7879; Fax: +39-011-670-7020.

**Abstract:** The detection of quantal exocytic events from neurons and neuroendocrine cells is a challenging task in neuroscience. One of the most promising platforms for the development of a new generation of biosensors is diamond, due to its biocompatibility, transparency and chemical inertness. Moreover, the electrical properties of diamond can be turned from a perfect insulator into a conductive material (resistivity $\sim$m$\Omega$ cm) by exploiting the metastable nature of this allotropic form of carbon. A 16-channels MEA (Multi Electrode Array) suitable for cell culture growing has been fabricated by means of ion implantation. A focused 1.2 MeV He$^+$ beam was scanned on a IIa single-crystal diamond sample (4.5$\times$4.5$\times$0.5 mm$^3$) to cause highly damaged sub-superficial structures that were defined with micrometric spatial resolution. After implantation, the sample was annealed. This process provides the conversion of the sub-superficial highly damaged regions to a graphitic phase embedded in a highly insulating diamond matrix. Thanks to a three-dimensional masking technique, the endpoints of the sub-superficial channels emerge in contact with the sample surface, therefore being available as sensing electrodes. Cyclic voltammetry and amperometry measurements of solutions with increasing concentrations of adrenaline were performed to characterize the biosensor sensitivity. The reported results demonstrate that this new type of biosensor is suitable for *in vitro* detection of catecholamine release.

---

## 1. Introduction

In recent years, an increasing interest in the understanding of the human brain function promoted the funding of international networks such as the Human Brain Project [1] in Europe and the BRAIN Initiative [2] in USA. A key mechanism of brain activity is the quantal release of bioactive molecules from neurons at the basis of synaptic transmission and hormone release. The detection of quantal exocytic events is therefore a prominent research field in neuroscience.

During exocytosis, membrane-bound vesicles storing transmitter molecules release their contents into the extracellular space. Many cell types share the secretory apparatus and its functionality, including the chromaffin cells of the adrenal gland, which release catecholamines (i.e. adrenaline, dopamine and noradrenaline) [3] and are widely used as a model system to study neuronal exocytosis. The quantal release of catecholamines can be assessed by amperometric trials, typically using carbon fiber microelectrodes (CFEs) [4]. Upon oxidization, each catecholamine molecule transfers two electrons to the polarized anode [4, 5]. The release from single vesicles is often preceded by a "foot" signal [6] associated with the opening and expansion of the fusion pore [7].

Standard CFE probes are characterized by chemical stability and biocompatibility, but can be hardly integrated into a miniaturized multi-electrode device, thus limiting their use in multiple single-cell recordings. Because of these shortcomings, other materials have been employed to produce "lab-on-a-chip" devices that are highly demanded in modern biotechnology to preserve living cells *in vitro* for long periods while revealing a broad range of electrical bio-signals. Such alternative technologies are based on silicon [8], gold [9, 10], platinum [11, 12], indium-doped tin oxide [13-15], diamond-like carbon [16-18], carbon nanotubes [19] and conductive polymers [20]. More recently, cellular biosensors based on nanocrystalline [21-27] and single-crystal [22, 28] diamond were developed with the purpose of overcoming the limitations of the above-mentioned materials in terms of transparency, bio-compatibility [29-39] and electrochemical window width [40].

In recent years, the Deep Ion Beam Lithography (DIBL) technique was optimized to microfabricate single-crystal diamond by means of scanning focused MeV ion microbeams. This approach allows to tune the optical properties of the material, both with regards to its refractive index [41-44] and to the formation of luminescent centers [45]. Also the structural properties of the material undergo significant modification upon MeV ion irradiation (surface swelling [46], stress induced effects [47-49]). Most importantly in the present context, by overcoming a critical fluence during diamond irradiation it was possible to create graphitic structures in single-crystal diamond. This approach takes advantage of the metastable nature of diamond, which can be converted into the stable allotropic form of carbon at room temperature and pressure conditions (i.e. graphite) by creating high defect concentration in the lattice [50-56].

Our previous studies on the fabrication of conductive graphitic microchannels in single-crystal diamond by DIBL led to the realization of a prototypical single cell biosensor. With such a device, the exocytotic activity from single chromaffin cells was measured by amperometry, showing the potential of further developing the diamond-based device into a multi cell sensor [57].

This work reports our progresses in the realization of multiple conductive micro-structures in single-crystal diamond and on the consequent electrochemical testing of a 16-channel multi electrode array (MEA) biosensor.

## 2. Experimental Section

A commercial synthetic single-crystal diamond grown by Chemical Vapor Deposition (CVD) was purchased from ElementSix (UK). The sample has size of 4.5×4.5×0.5 mm$^3$ and is classified as type IIa ("optical grade") with substitutional nitrogen and boron concentrations lower than 1 ppm and 0.05 ppm, respectively. The diamond is cut along the [100] crystal direction and is optically polished on the two opposite large faces.

The sample was implanted across one of the polished surfaces with 1.2 MeV He$^+$ ions at the AN2000 accelerator of the INFN National Laboratories of Legnaro (INFN-LNL) [58] with a focused ion beam, in order to deliver an ion fluence of $1\times10^{17}$ cm$^{-2}$. The microbeam spot was ~10 μm in diameter. All implantations were performed at room temperature using currents lower than 30 nA to minimize sample heating.

MeV ions induce structural damage in matter mainly at their end of range, where the cross section for nuclear collisions is strongly enhanced, after the ion energy is progressively reduced by Coulombian interactions occurring in the initial stages of the ion path. Figure 1 shows the strongly non-uniform depth profile of the damage density evaluated in first approximation as the product between the linear damage density (number of vacancies per unit length per incoming ion) evaluated with Monte Carlo numerical simulations and the implantation fluence (number of implanted ions per unit surface). The numerical Monte Carlo code employed in this work is "Stopping and Range of Ions in Matter" (SRIM), in its 2013.00 version [59]. In the simulation, the "detailed calculation with full damage cascade" mode was adopted, while taking an atom displacement energy value of 50 eV [60, 61]. The high density of damage induced by ion implantation promotes the conversion of the diamond lattice into an amorphous phase within a layer ~250 nm thick ~~which is~~ located ~2 μm below the sample surface.

**Figure 1** Left: SRIM evaluation of vacancy density profiles generated in diamond by 1.2 MeV He$^+$ irradiation after crossing a copper mask with increasing thicknesses. The graphitization threshold ($9\times10^{22}$ cm$^{-3}$) is indicated with a dashed line. Right: schematics of the variable-thickness mask configuration, which allows the definition of the sub-superficial highly damaged layer at variable depth.

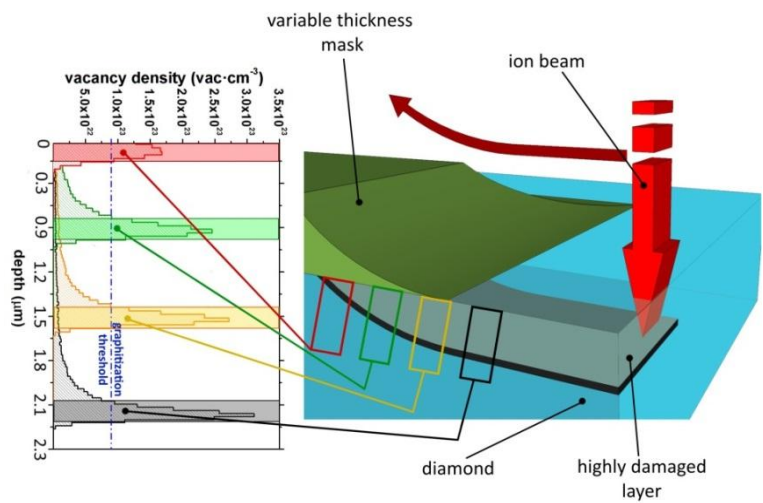

In order to connect the amorphized structures to the sample surface, a three-dimensional masking technique was employed to modulate with high spatial accuracy the penetration depth of the ions from their range in the unmasked material up to the sample surface with increasing thickness of stopping material [56], as schematically shown in Figure 1.

A smooth mask profile is fundamental for this scope, because the channels must be gradually bent towards the surface; a too sharp profile would lead to a channel disconnected from the surface. Variable thickness masks were obtained by metal evaporation through suitable hole pattern realized by laser cutting of a thin aluminum foil. The foil was then placed at a distance of ~300 μm from the diamond surface: in this configuration, a "shadow" effect determines deposited masks characterized by a constant thickness in their central regions and with smooth and slowly degrading edges at their sides.

Ion implantation was performed by scanning an ion beam along a linear path having both endpoints in correspondence of the above-mentioned metal masks. As the beam scan progresses, incident ions cross the masks in correspondence of an increasing thickness of metal, thus progressively reducing their range in the diamond layer and emerging up to the surface [56]. As an example, Figure 1 shows the vacancy profiles generated by 1.2 MeV $He^+$ ions in diamond after crossing increasing thicknesses of copper.

This procedure was repeated 16 times in a suitable geometry, thus allowing the fabrication of 16 micro-channels. As shown in Figure 2, for each channel one emerging endpoint was located in the central area acting as a bio-sensing electrode for cellular *in vitro* recording, while the other emerging endpoint was located in a peripheral region of the device, thus providing a contact for chip bonding onto the carrier and towards the front-end electronic board. The geometry of the device was defined to have the central sensing electrodes placed as a 4×4 matrix with an inter-electrode spacing of 200 μm.

**Figure 2** Top-view optical micrograph of the diamond-based biosensor: the 16 graphitic channels are ~20 μm wide, ~1 mm long, ~250 nm thick, with sensing areas of ~20×5 $\mu m^2$. The diamond substrate is mounted and wire-bonded on the chip carrier.

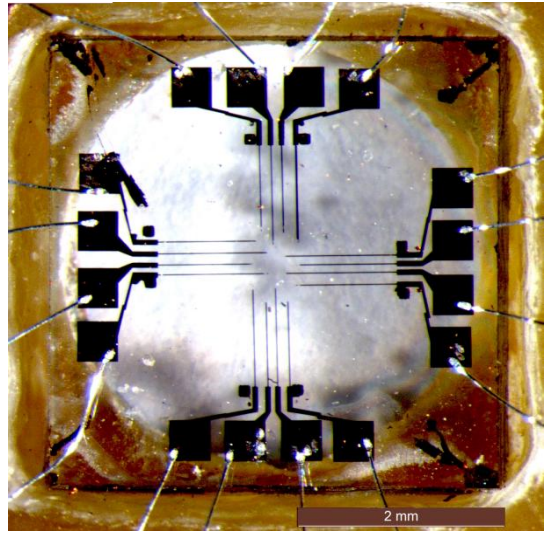

After ion implantation, the Cu masks were removed from the surface. The sample was then annealed in high vacuum conditions (~$10^{-6}$ mbar) at 900 °C for 1 h, to convert the highly-damaged regions located at the ion end of range to a graphitic phase while removing the structural sub-threshold damage introduced in the layer overlying the damaged region.

The electrical characterization of the channels consisted of two-terminal current-voltage (I-V) measurements and was performed with a system of microprobes operating at room temperature. The voltage sweep was performed between -3 V and +3 V. The maximum sensitivity of the electrometer is 10 fA and the accuracy is typically better than 1%, depending on the selected range and quality of the contacts. In order to improve the quality of the electrical contact between the channels and the microtips, temporary 200 nm thick Ag contacts were deposited in correspondence of the emerging endpoints of the graphitic micro-structures.

After these preliminary tests, the Ag contacts were removed and the final device was assembled by fixing the diamond sample into the chip carrier. Metal contacts defined by means of electron beam lithography were deposited over the peripheral emerging endpoints of the channels: 15 nm thick chromium and 150 nm thick copper layers were successively evaporated. The chromium buffer layer plays a crucial role in the improvement of the adhesion between the Cu layer and the diamond surface, since after a heating process at 500 °C for 1 hour it forms a conductive chromium carbide film. The above-mentioned metal electrodes were connected with the chip carrier employing a conventional wire-microbonder. Both the Cr/Cu electrodes and the bonding wires were then passivated with sylgard, in order to electrically insulate them with a bio-compatible plastic from the physiological solution which will fill the perfusion chamber during the *in vitro* tests (see below for more details).

The chip carrier was designed to perform *in vitro* measurements on cells directly cultured on top of the diamond, and was therefore equipped with a 60 ml perfusion chamber. The experimental setup was completed by the front-end electronics, where the chip carrier is directly plugged-in, an A/D converter unit (National Instruments USB-6216) and acquisition software developed in LabView. The front-end electronic components, expressly designed for amperometric measurements, are based on low-noise transimpedance amplifiers, having an input bias current of 1 pA and the gain set by feedback-resistors of 100 MΩ. Subsequently, the signals are sent via differential lines to the ADC unit, which is equipped with a tailored interface daughterboard including differential receivers followed by Bessel low-pass filters of the sixth order with a bandwidth of 1 kHz. The 16 bit A/D converter runs at a sampling rate of 4 kHz per channel and is ultimately connected to the computer over a hi-speed USB link. The acquisition software allows, by means of the graphical user interface, both cyclic voltammetry and chronoamperometry measurements with full control of all parameters like biasing voltage, scan rate, ADC voltage range, sampling rate etc.

Different solutions were used for cyclic voltammetry trials: a standard saline solution (Tyrode) containing (in mM): 128 NaCl, 2 $MgCl_2$, 10 glucose, 10 HEPES, 10 $CaCl_2$ and 4 KCl. Increasing doses of adrenaline (10, 50 and 100 μM) or dopamine (100 μM) were then added to the Tyrode solution with the purpose of testing the detection of the oxidation peak of the catecholamines.

## 3. Results and Discussion

The electrical properties of the 16 micro-channels were investigated by I-V measurements. I-V curves were obtained by probing two contacts located at the endpoints of each single channel, as shown in Figure 3. The I-V characteristics show an ohmic behavior and a resistance of ~5–10 kΩ was obtained, except for channel #16 that exhibited no electrical conductivity. We attributed the latter feature to an accidental failure during the DIBL fabrication of the channel, which determined its discontinuity and/or an incomplete emergence to the sample surface.

**Figure 3**. (**a**) 3D schematics of the diamond-based biosensor configuration during I-V measurements. (**b**) I-V characteristics of the 16 graphitic channels.

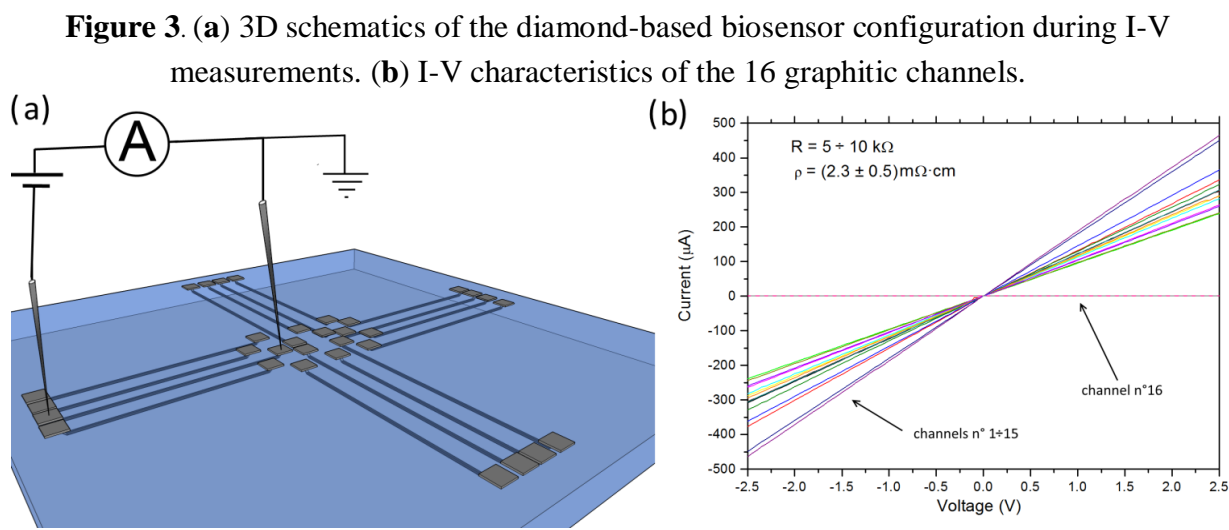

The resistivity of each conductive channel was estimated from its resistivity and geometrical dimensions. In particular, the lateral dimensions (length ~950–1200 μm, width ~10-15 μm) were measured by optical microscopy, while the thickness (~250 nm) was estimated from the intersection of the graphitization threshold with the SRIM-derived vacancy density profile (see Figure 1). With the exception of the above-mentioned channel #16, the obtained mean value of the resistivity of the channels was comparable with that of common polycrystalline graphite, i.e. $\rho = (2.3 \pm 0.5)$ mΩ cm compared to ~1.3 mΩ cm [62]. It is worth mentioning that the channels were embedded in the fully insulating ($\rho > 1$ TΩ) single-crystal diamond matrix.

With the aim of investigating the dark current and the electrochemical response of the device in working condition, the redox properties of the above-mentioned saline solution (Tyrode) and the oxidation potential of adrenaline and dopamine were investigated by means of cyclic voltametric measurements.

A triangular voltage waveform with a scan rate of 20 mV s$^{-1}$ and ranging from 0 V to +1.2 V was applied to the graphitic micro-electrodes with respect to the Ag/AgCl quasi-reference electrode. The voltage-dependent redox currents collected at the micro-electrodes were recorded and processed.

The device was perfused with a standard Tyrode solution (see the previous section for details) and no redox activity was detected within the anodic range of the hydrolysis window, showing a leakage current lower than 10 pA up to a bias voltage of +900 mV. Subsequently, the electrochemical response of the device was first tested in the presence of increasing concentrations of adrenaline (10, 50, 100 μM and 1 mM) and then in the presence of dopamine (100 μM). Figure 4 shows the response of one representative electrode (channel #4) to the different assayed solutions. It has to be mentioned that even if measurements performed with dopamine induce electrodes' fouling [63], our diamond-based biosensor can be easily cleaned guaranteeing the restoration of the previous performances.

**Figure 4** Cyclic voltammetric scans at 20 mV s$^{-1}$ rate of a 0–1.2 V voltage applied to the #4 sensing microelectrode with respect to the quasi-reference Ag/AgCl electrode in the presence of Tyrode buffer (**a**) and Tyrode spiked with: 10 μM adrenaline (**b**), 50 μM adrenaline (**c**), 100 μM adrenaline (**d**), 1 mM adrenaline (**e**) and 100 μM dopamine (**f**).

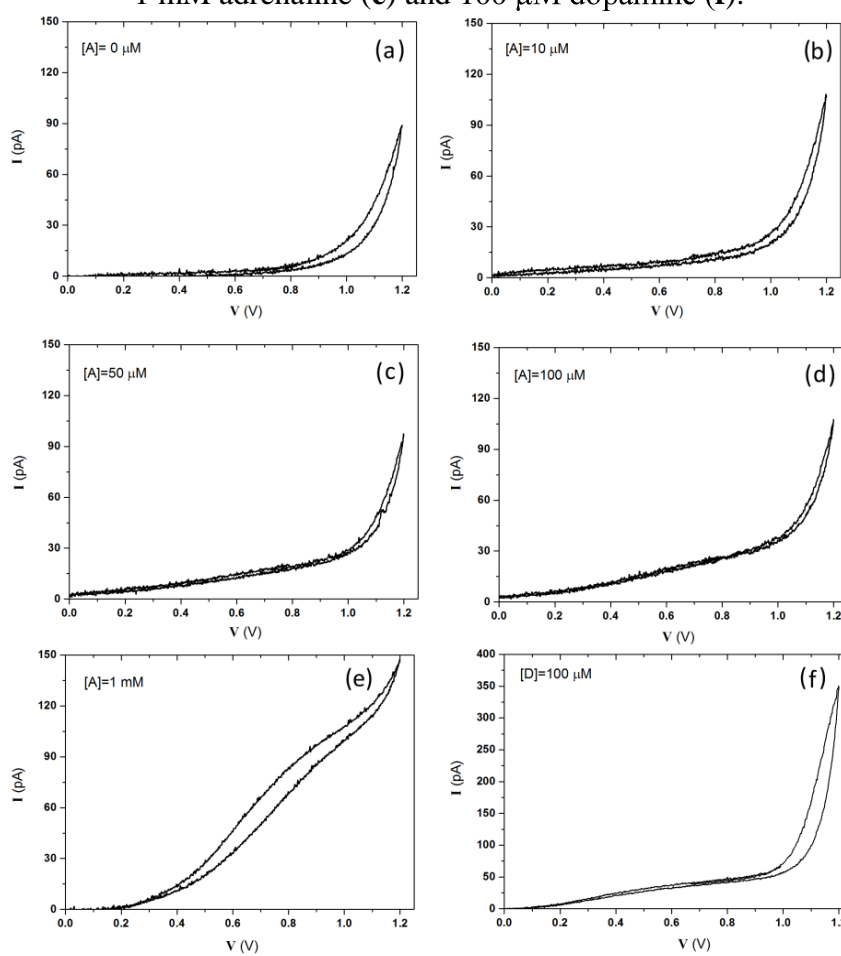

These tests confirmed the possibility of employing our diamond-based sensor for electrochemical measurements thanks to the overpotential for oxygen-evolution occurring around +1 V at pH 7.4 (Tyrode buffer) in the anodic range. This peculiarity is crucial for the detection of the oxidation current of the more common neurotransmitters, which were observed between +650 mV and +850 mV. Moreover, the low leakage currents ensure a limit of detection which allows the recording of spike signals (> 20 pA) produced by the exocytic events in neuroendocrine cells [57].

The curves related to the adrenaline allowed the evaluation of the optimal bias voltage for the subsequent amperometric measurements and for the foreseen *in vitro* measurements, which was set to +800 mV, corresponding to the maximum value of the (oxidation current):(water hydrolysis) ratio, i.e curve B, C, D and E vs. A, respectively (see Fig. 4).

With the purpose of quantifying the sensitivity of the device to adrenaline, the oxidation potential was set to the above-mentioned optimal value (i.e. V = +800 mV) during chronoamperometric tests. Figure 5A shows recordings of one representative electrode (channel #4) in response to 1÷100 μM adrenaline perfusions. With the exception of the above-mentioned channel #16, the response of all working electrodes was consistent with what reported in Fig. 5a. For each concentration, data were acquired for 1 min and then the chip was washed with Tyrode solution and subsequently tested with a different concentration. Averaged data were fitted by linear regression ($r = 0.99$, slope $(2.2 \pm 0.1)$ pA μM$^{-1}$), indicating a direct proportionality between the adrenaline concentration in the solution and the amperometric response at V = +800 mV of all micro-electrodes, as shown in Figure 5B. It is worth remarking that such proportionality can be employed in the field of analytical chemistry, ensuring the quantification of unknown concentrations of catecholamines in solution.

**Figure 5** (**a**) amperometric chronograms in correspondence of the reported adrenaline concentrations, polarized to +800 mV with respect to the quasi-reference electrode. For each adrenaline concentration, the acquisition lasted 60 s, while in the graph only 1 s of the recording is reported. Consecutive recordings using increasing concentrations of adrenaline were separated by time breaks, in order to properly remove the previous solution and wash the device. (**b**): 60-second average amperometric signal recorded at +800 mV for a representative electrode (channel #4) as a function of adrenaline concentration; experimental data are shown in circular dots, while the linear fit is reported in continuous line.

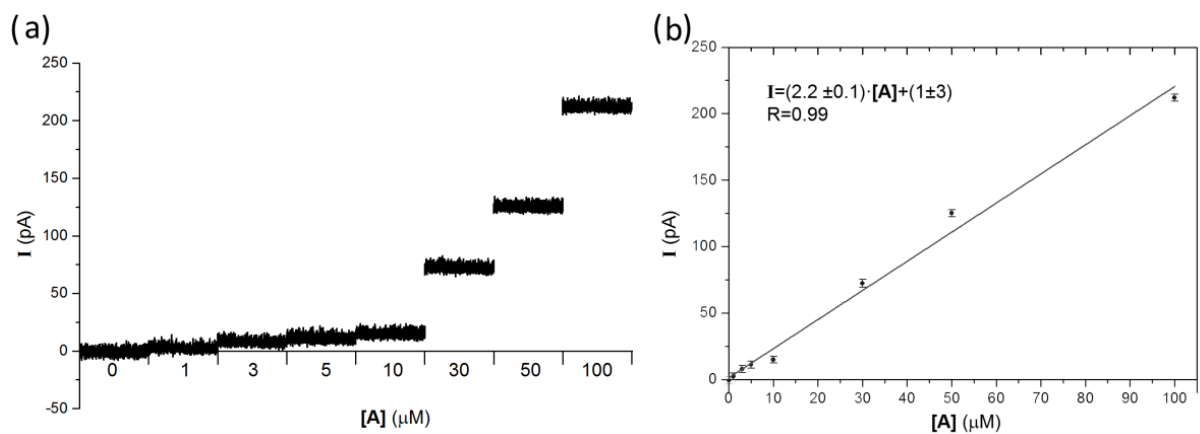

## 4. Conclusions

The fabrication process of a single-crystal-diamond-based sensor and its systematic electrical and electrochemical characterization has been reported. 16 graphitic sub-superficial micro-channels were realized by means of Deep Ion Beam Lithography (DIBL) in a diamond matrix offering the possibility of recording simultaneous electrochemical signals. High sensitivity and linear response to adrenaline concentration have been demonstrated, and a wide potential margin from the oxygen-evolution guarantees several physiological applications of the device, since different catecholamines can be detected. The presented biosensor offers the possibility of simultaneously investigating *in vitro* exocytotic events from multiple excitable cells cultured directly on a bio-compatible, transparent and robust platform. This aspect will be systematically investigated in future studies.

## Acknowledgments

The authors wish to thank Giampero Amato, Luca Boarino, Antonio Zampieri and Chiara Portesi for the kind support. This work is supported by the following projects: "DiNaMo" (young researcher grant, project n° 157660) by National Institute of Nuclear Physics; FIRB "Futuro in Ricerca 2010" (CUP code: D11J11000450001) funded by MIUR, "A.Di.N-Tech." (CUP code: D15E13000130003) and "Linea 1A - ORTO11RRT5" funded by the University of Torino and "Compagnia di San Paolo". "Nanofacility Piemonte" is a laboratory at INRiM supported by the "Compagnia di San Paolo" Foundation. The MeV ion beam lithography activity was performed within the "Dia.Fab." experiment of the INFN Legnaro National Laboratories.

## Author Contributions

F.P, A.B. and J.F prepared the sample. S.G and V.C. did the experiments. A.P. realized the acquisition hardware and software. L.C. did the metal pad deposition. E.E. did the electron beam lithography. P.O. and E.C. designed the experiment. F.P and A.B. wrote the manuscript with input from all coauthors. All authors reviewed the manuscript.

## Conflicts of Interest

The authors declare no conflict of interest.